# Strong pretransitional anomaly near glass transition

Szymon Starzonek*, Aleksandra Drozd-Rzoska, Sylwester J. Rzoska

*Institute of High Pressure Physics of the Polish Academy of Sciences, Warsaw, Poland*

*The Author to whom correspondence should be addressed: starzoneks@unipress.waw.pl

## ABSTRACT

In this report we present the first ever results of strong pretransitional effect observed for the maximum of absorption peak $\varepsilon''_{max}(T)$ in supercooled liquid-crystalline systems near the glass transition temperature. This anomaly may be described by critical-like relation with critical exponent $\alpha = 0.5$, which corresponds well with previous results for static electric permittivity $\varepsilon_s(T)$. Such a behaviour may suggest thermodynamic character of glass transition.

The description of the glass transition as a typical phase transition is still a subject of academic debate [1-4]. On the one hand, it is believed that the glass transition is a continuous transition, which begins at the temperature $T_g$ and ends only around the Kauzmann temperature. There is still insufficient experimental evidence to unambiguously define its character. Colby and Erwin found that the dynamics of the transition to the glass state is related to the critical point ($T_c$) below $T_g$ [5,6]. Then Tanaka [7] created a model in which he assumed that in each liquid there is a tendency to local aggregation of the bond-ordering and / or density driven type, as a result of which two kinds of local order parameters can always be defined. Depending on the degree of dominance of one of the aggregation types and their mutual correlation in a given liquid, the Tanaka model leads to the classic liquid-gas critical point, glass transition or liquid-liquid transition in a one-component system. In the last year, several studies were published showing the existence of the liquid-liquid transition near the temperature $T_g$ in metallic glasses [8].

All this means that the discussion on the critically similar description of the vitreous dynamics returns constantly. It should be emphasized that the relationship $\varepsilon_s(T)$ and $\varepsilon''_{max}(T)$ for the glass transition in any system has not been presented so far. Therefore, when analyzing supercooled liquid crystals, it was decided to pay attention to the temperature dependence of the dielectric constant or the maximum absorption peak in the vicinity of the glass transition temperature. This work is the first attempt to address the above issue.

According to the previously published results of detection of phase transitions in liquid crystals, one should expect the characteristic behavior of the dielectric constant when approaching the transition temperature, which has been called the pretransitional anomaly. This phenomenon consists in the cancellation of permanent dipole moments through the antiparallel arrangement of molecules inside the critical fluctuations. The pretransitional anomaly can be described by the critical-like equation with the exponent $\alpha = 0.5$ [1-4]:

$$\varepsilon_s(T) = \varepsilon^* + a(T - T^*) + A(T - T^*)^{1-\alpha}, T > T_c,$$

(1)

where $a, A$ are amplitudes, $\alpha = 0.5$ is the critical exponent, and $(T^*, \varepsilon^*)$ is the point of an ideal continuous phase transition below $T_c$. It is worth mentioning that for nematic liquid crystals $\Delta T_c = T_c - T^* = 1 - 2$ K [9-11].

The description and analysis of the pretransitional anomaly is strongly dependent on the type of thermodynamic variables (field variables, i.e. temperature, pressure, volume, density, molar concentration), as well as the direction of approaching the critical point.

In the case of the tested liquid crystals, the existence of a pretransitional anomaly in the vicinity of the glass transition temperature for the dielectric constant $\varepsilon_s(T)$ was not demonstrated (Fig. 1A). It is directly related to experimental limitations, because near $T_g$ the static area of the real part is in the low frequency range ($f < 10^{-3}$ Hz), which extends the measurement time to several days for one substance. However, the analysis of the $\varepsilon''_{max}(T)$ value, which is the maximum of the absorption curve, has proven effective and useful, showing a strong pretransitional effect in the supercooled phase just before the temperature $T_g$. A similar effect has previously been found in another liquid crystal from the group of n-cyanobiphenyls (4CB) for an isotropic-nematic liquid transition [5]. The pretransitional anomaly $\varepsilon''_{max}(T)$ can be described analogously to (1):

$$\varepsilon''_{max}(T) = \varepsilon''^{*}_{max}(T) + a'(T - T^*) + A'(T - T^*)^{1-\alpha}, T > T_c.$$

*(2)*

From the physical point of view, it is important to interpret the maximum value of the absorption peak, which we understand as energy dispersed in the system. In order to facilitate the analysis, it is worthwhile to differentiate the equation (55), which gives us Fig. 1B illustrates with $\varepsilon''_{max}(T)$ in the vicinity of $T_g$ for the tested liquid crystals (E7, 5*CB, 8*OCB), and Table 1 gives the fit parameters by equation (2)

$$\frac{\partial \varepsilon''_{max}(T)}{\partial T} = a' - A'(1 - \alpha)(T - T^*)^{-\alpha}$$

*(3)*

Table 1. Parameters fitted by Eq. (2).

| System | $T_g$ (K) | $T^*$ (K) | $\varepsilon_{max}''^*$ | $a$ | $A$ | $\alpha$ |
|---|---|---|---|---|---|---|
| E7 | 210 | 204 | 1.65 | -0.15278 | 2.063 | 0.5 |
| 8*OCB | 223 | 221 | 3.65 | -0.05009 | 0.5734 | 0.5 |
| 5*CB | 216 | 215 | 1.76 | -0.2141 | 1.20706 | 0.5 |

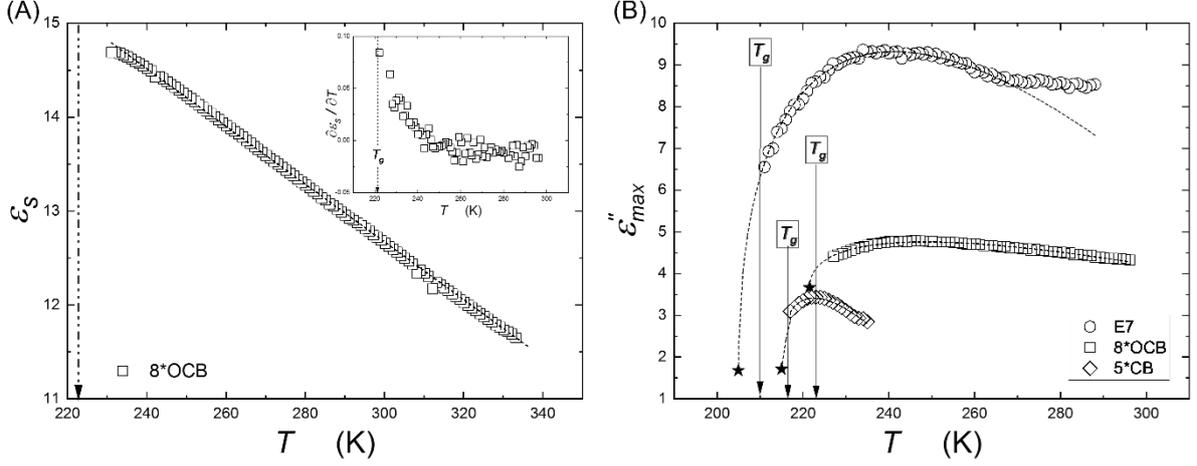

FIG. 1 (A) Static dielectric permittivity of 8*OCB liquid crystal in isotropic phase near $T_g$. (B) Pretransitional anomaly in the vicinity of glass transition for all studied system. Dashed lines denote relation (2). Fitting parameters are given in Table 1.

Derivative-based analysis of $\varepsilon_{max}''(T)$ greatly facilitates the adjustment of the pseudocritical function to the experimental data due to the sensitivity to any unnoticeable distortions. Moreover, due to the equation (2), the determination of the temperature $T^*$ becomes a simple matter from the mathematical point of view.

Fig. 2 illustrates the differential analysis of the critical-like equation describing the maximum of the absorption curve $\varepsilon_{max}''(T)$ together with the fitted equation (2). From Figs. 1 and 2, the existence of a strong pretransitional anomaly in the vicinity of the glass transition temperature $T_g$ directly follows. It is related to the increased energy dissipation in the system and the broadening of the absorption peak and suggests the presence of strong heterogeneities in the dynamic description. On the other hand, when analysing the glass transition on the basis of critical phenomena, a strong pretransitional anomaly indicates the occurrence of fluctuations near $T_g$ and allows to treat the glass transition as a kind of phase transition [6,12,13].

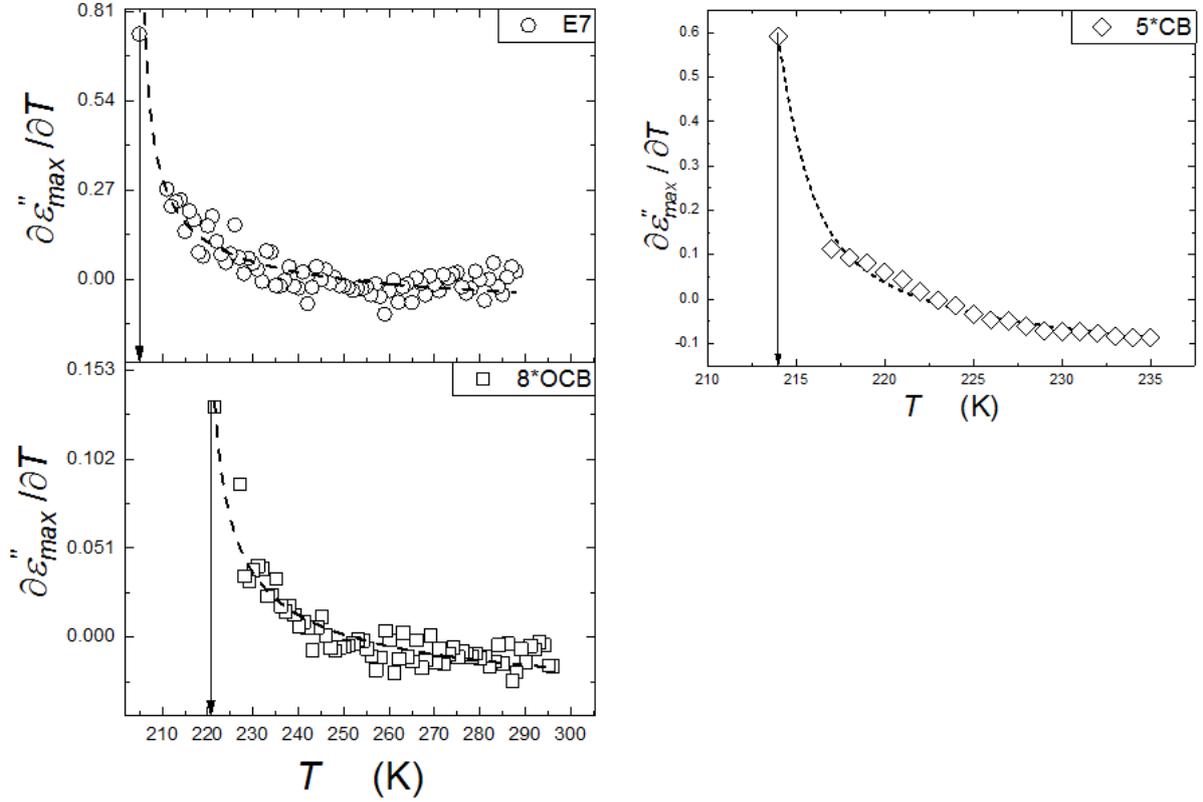

FIG. 2 Derivative-based analysis by Eq. (3) of eps max for studied system. Strong anomaly occurs near glass transition temperature.

Taking into account that the temperature of the ideal continuous phase transition $T^*$ is located in the area between the glass transition temperature $T_g$ and the Kazuzmann temperature, it can be assumed that it is at this temperature that the dynamics of the system changes, manifesting itself in the transition from non-Arrhenius to Arrhenius description ($\Delta E_a(T) \to \Delta E_a = const$) [1-6,11,12]. The above change of dynamics character should be related to the hypothetical relaxation time in $T_g$ equal to 100 s, which was previously assumed as the glass transition point. However, the experimental results so far show that the glass transition, manifested by $\Delta E_a(T) \to \Delta E_a = const$, takes place at a temperature lower than $T_g$, which suggests its blurred character [1-6,11,12]. On the basis of the data obtained in this paper, as well as the previously published papers, it can be proposed that the transition to the state of glass is continuous [1-7,11,12]. From a practical point of view, it can be assumed that the experimentally determined temperature $T_g$ is the beginning of the transition, while the Kauzmann temperature is the point of reaching thermodynamic equilibrium [28, 32, 84-86].

Concluding, the temperature dependence of $\varepsilon''_{max}(T)$ shows a pretransitional anomaly with the critical exponent $\alpha = 0.5$ for all liquid crystals, which suggests the existence of a phase transition around the glass transition temperature $T_g$. The latter may suggest thermodynamic character of the glass transition.